# NIMO: A Software Platform for Closed-Loop Materials Exploration with Diverse AI Algorithms

**Ryo Tamura[*1,2], Naruki Yoshikawa[1], Koji Tsuda[2,1], and Shoichi Matsuda[*3]**

*1 Center for Basic Research on Materials, National Institute for Materials Science, 1-1 Namiki, Tsukuba, Ibaraki 305-0044, Japan*

*2 Graduate School of Frontier Sciences, The University of Tokyo, 5-1-5 Kashiwa-no-ha, Kashiwa, Chiba 277-8561, Japan*

*3 Center for Green Research on Energy and Environmental Materials, National Institute for Material Science, 1-1 Namiki, Tsukuba, Ibaraki 305-0044, Japan*

## Abstract

Self-driving laboratories (SDLs), where artificial intelligence proposes subsequent experiments and robotic systems execute them, are rapidly becoming the vanguard of materials discovery. A critical bottleneck, however, lies in seamlessly bridging diverse AI algorithms tailored for specific exploration goals with the heterogeneous robotic hardware found across different laboratories. Here, we present NIMO, an open-source software platform designed to dissolve this barrier through three core paradigms: a modular AI–robot decoupling mediated via simple CSV file exchange, a discrete candidate-pool architecture that seamlessly absorbs domain knowledge, and a unified Python interface pre-loaded with twelve distinct AI algorithms. In this Perspective, we review the operational principles of each algorithm alongside six diverse SDL implementations driven by NIMO, covering electrolyte discovery, organic synthesis, thin-film exploration, fuel-cell process informatics, coffee-ring phase exploration, and legacy liquid-handling automation. One of these also demonstrates NIMO's seamless interoperability with the IvoryOS orchestration framework. To democratize autonomous science, we also introduce a no-code desktop application that enables intuitive, human-in-the-loop exploration for non-programmers. NIMO is freely available at https://github.com/NIMS-DA/nimo, offering a versatile, plug-and-play foundation to accelerate autonomous materials exploration across diverse experimental landscapes.

## 1. Introduction

The discovery of advanced materials drives technological progress across a wide range of fields, from clean energy to next-generation spintronics. However, the pace of materials discovery based on conventional trial-and-error experimentation has become a major bottleneck, falling increasingly behind societal and industrial demand. To address this gap, autonomous materials exploration has emerged as a powerful new paradigm, in which artificial intelligence (AI) takes on the role of experimental planner and selects the next experimental conditions on the basis of accumulated data. Even when the experiments are carried out by human researchers, AI-guided planning has been shown to outperform conventional experimental design strategies by a large margin[1,2]. When the experimental step is also handed over to robotic platforms, this paradigm goes one step further. This combination of autonomous planning and automated experimentation defines the self-driving laboratory (SDL)[3–7]. In an SDL, the AI proposes the next experiment, a robotic system performs it without human intervention, and the resulting data are fed back to the AI to guide the next decision, thereby closing a feedback loop running uninterrupted for extended period. Building such a closed loop requires the seamless integration of diverse instruments, robotic hardware, and AI algorithms into a single coherent system. Orchestration software (OS) sits at the center of this integration, coordinating the flow of commands and data among the individual modules and managing the overall experimental workflow. Reflecting its central role, a growing number of open-source

OS frameworks designed for SDLs have been developed in recent years [8–20].

Among the AI algorithms used in SDLs, machine-learning-based black-box optimization (BBO) has become the dominant approach[21]. Particularly within this context, Bayesian optimization (BO) is widely utilized to guide the search toward optimal experimental conditions. A comprehensive survey of available BO libraries is detailed in the SDL review[4]. A notable example is ChemOS, which integrates Phoenics[22], a BO framework tailored for chemistry. However, experimental objectives in materials science are highly diverse, often requiring tailored exploration paradigms. In some cases, the goal is to identify the single best material; in others, to find a set of materials whose properties fall within a target range; and in others, to obtain a diverse set that broadly covers the property space. Addressing this diversity of goals requires a corresponding various of algorithms, each tailored to a specific objective. To meet this need, we have developed NIMO (originally NIMS-OS, the NIMS Orchestration System), an open-source library[23]. NIMO features a built-in collection of AI algorithms that encompass not only BO but also alternative search methods, covering a wide range of materials exploration scenarios. Whereas other open-source OSs typically focus on workflow orchestration and rely on external BO packages for the AI layer, NIMO directly embeds a wide variety of AI algorithms within its framework. NIMO has two further distinctive aspects. First, the interface between the AI algorithms and the experimental workflow is handled through a simple standard format based on CSV file exchange (see Fig. 1). Second, NIMO operates on a discrete candidate pool: the user enumerates all physically realizable experimental conditions in a single *candidates.csv* file, from which the AI selects the next experiment.

These design choices benefit both fully automated SDLs and human-in-the-loop exploration. For SDLs, NIMO greatly reduces the engineering effort required for integration. Connecting NIMO to a robotic system requires writing only a single interface module that links the robot to NIMO's CSV-based input/output. Once this single connection is prepared, all of NIMO's built-in AI algorithms become immediately available, eliminating the need to develop a separate integration for each algorithm. The CSV-based interface also allows NIMO to connect to robotic systems regardless of their underlying programming language (Python, Visual Basic, LabVIEW, or vendor-specific scripting languages). Furthermore, it enables integration with other OSs, allowing NIMO to serve as the AI decision engine within those systems. In addition, the candidate-pool design prevents the AI from proposing infeasible experiments, thereby avoiding experimental failures in closed-loop campaigns. For human-in-the-loop exploration, NIMO can serve as a stand-alone solution: a single package provides access to many algorithms, and the closed loop is realized simply by reading the proposed conditions from a CSV file, performing the experiments, and writing the results back into the same file.

NIMO is distributed as a Python package on PyPI (*pip install nimo*), with source code hosted at https://github.com/NIMS-DA/nimo. It also ships with a desktop application for users without programming experience. The official documentation at https://nims-da.github.io/nimo/en/ provides full API references and worked examples. This Perspective provides a comprehensive overview of NIMO, including its architecture, algorithms, and applications. Section 2 describes the closed-loop architecture and the file-exchange protocol that makes interoperability possible. Section 3 details the design and construction of the candidate pool, a key source of NIMO's flexibility. Section 4 surveys the twelve AI algorithms currently implemented, organized by exploration goal. Section 5 presents six practical SDL implementations across diverse domains: electrolyte discovery, organic synthesis, thin-film exploration, fuel-cell process informatics, coffee-ring phase exploration, and legacy liquid-handling automation. Section 6 introduces NIMO Desktop, a no-code GUI that makes NIMO accessible to non-programmers. Section 7 concludes with our roadmap, including continuous expansion of the algorithm portfolio and the path toward a globally connected SDL network.

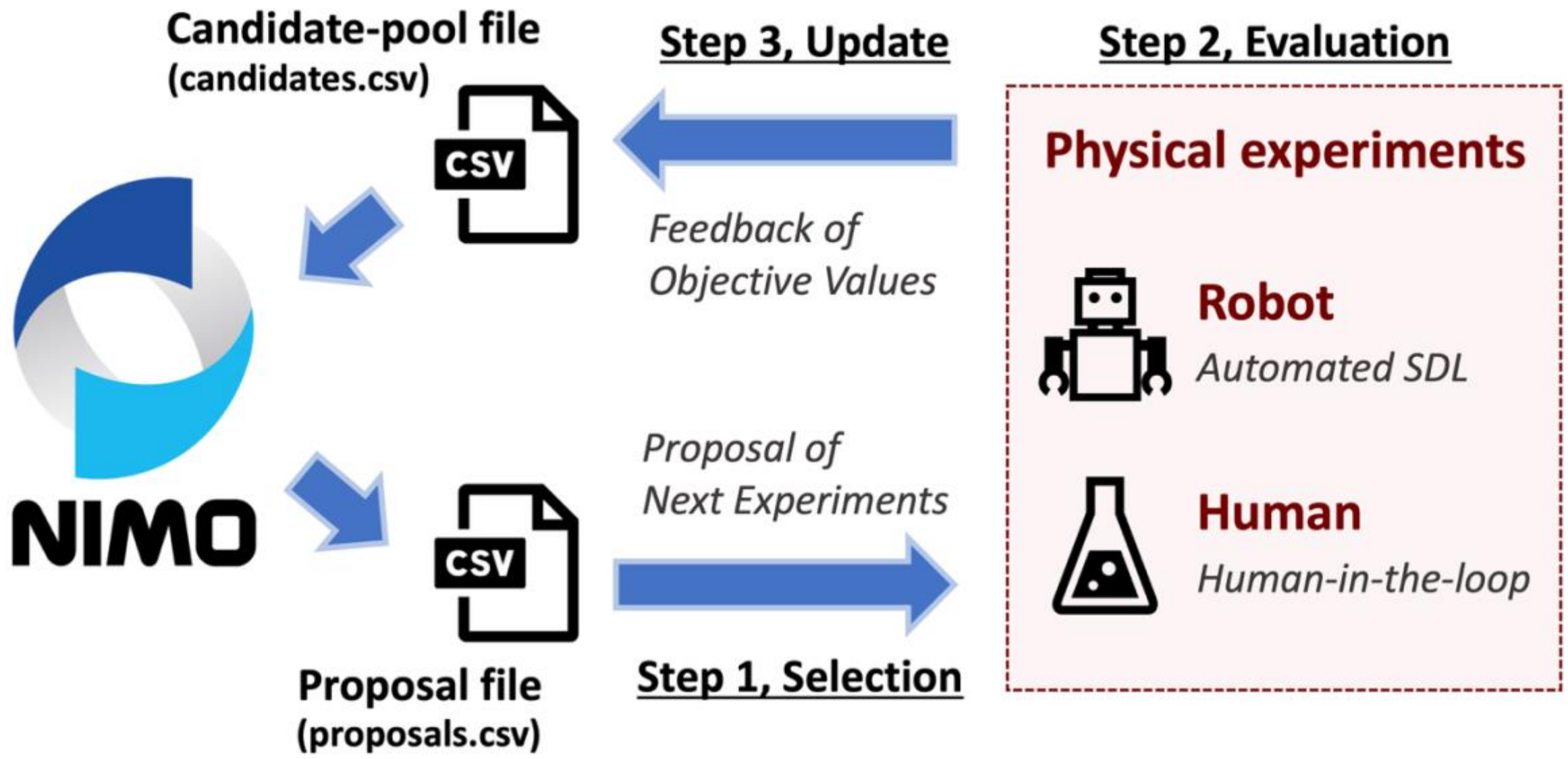

**Fig. 1** Conceptual diagram of the NIMO-enabled closed-loop workflow. NIMO and physical experiments communicate exclusively through two CSV files. In Step 1 (Selection), NIMO selects the next experimental conditions from the candidate-pool file (*candidates.csv*) and writes them to the proposal file (*proposals.csv*). In Step 2 (Evaluation), the proposed experiments are performed by either a robotic system or a human operator. Finally, in Step 3 (Update), the resulting objective values are written back to update the candidate-pool file (*candidates.csv*), restarting the cycle.

## 2. Closed-Loop Architecture

NIMO realizes the fully autonomous experimental loop illustrated in Fig. 1 through an iterative excecution cycle between NIMO and experimental evaluations:

**Step 1, Selection.** An AI module in NIMO reads a candidate-pool file (CSV) describing the materials search space (detailed in Sec.3), selects promising experimental conditions, and outputs them to a proposal file (CSV). This step is driven by the *nimo.selection* function.

**Step 2, Evaluation.** A robotic system or a human operator reads the proposal file and executes the corresponding experiments.

**Step 3, Update.** The candidate-pool file is updated with the generated experimental results, and control returns to Step 1. This update process is handled by the *nimo.output_update* function.

The decisive design choice of NIMO is that these three steps communicate exclusively via physical files rather than in-memory Python objects. Although file-based exchange might initially appear less elegant than direct, in-memory API calls, it offers profound practical advantages. Because the inter-module interface is restricted to plain CSV files, the AI module remains entirely agnostic to the downstream software controlling the experiments; conversely, the experimental hardware layer operates without needing to know that the AI algorithm is executed in Python. Consequently, any platform capable of reading and writing CSV files, be it Python, MATLAB, LabVIEW, Visual Basic, a shell script, or even a human manually entering data, functions as a valid, plug-and-play module.

Furthermore, this CSV-based interface enables NIMO to serve as a universal decision-making engine for external orchestration software (OS) platforms. This operational paradigm allows users to leverage the advanced workflow-design features of other OSs while seamlessly harnessing the diverse suite of experimental design algorithms embedded within NIMO. We have demonstrated this approach through a recent integration with IvoryOS[24], an OS that provides drag-and-drop web interfaces for Python-based SDLs[20]. By combining NIMO's AI planners with IvoryOS's device-control and workflow-management capabilities, users can construct significantly richer SDL ecosystems than either system can support alone.

## 3. Candidate Pool Design

A defining feature of NIMO is its reliance on a discrete candidate pool. In conventional BBO methods, the search space is typically defined as a continuous bounded box of parameter ranges, leaving the optimizer free to propose any coordinate within those boundaries. Consequently, the burden falls on the user to determine whether a given proposal is physically viable. NIMO inverts this paradigm: the user pre-enumerates every experimentally feasible condition, and the AI is strictly constrained to select from this predefined list. As a result, the AI never proposes an infeasible experiment without requiring explicit algorithmic constraints, which substantially bolsters operational reliability in fully autonomous automated workflows. This candidate-pool approach yields three practical advantages:

**Prior knowledge integration:** Domain expertise can be embedded directly into the search space during pool construction. For instance, a user can dictate that one parameter must always exceed another, or that a highly sensitive parameter should be sampled at a finer resolution. This eliminates the need to formulate such heuristic rules post-hoc as penalty terms or algorithmic constraints on the AI side.

**Flexible constraints:** The same mechanism effortlessly accommodates complex experimental rules that are notoriously difficult to formulate mathematically in continuous search spaces. For example, composition constraints, such as requiring three fractional components to sum to exactly 100%, as well as equipment-imposed physical restrictions, can be strictly enforced simply by omitting invalid conditions from the candidate list.

**Mixed-variable support:** Continuous, discrete, and categorical variables can coexist seamlessly within the same candidate pool. In standard BBO frameworks, combining heterogeneous data types often requires complex internal encoding schemes or specialized kernels that restrict algorithm selection. In NIMO, because the search space is pre-materialized as a discrete tabular dataset (such as a CSV file), every experimental candidate is simply treated as a single row. Consequently, any combination of variable types, including numeric continuous parameters, discrete process steps, and categorical factors represented as integers or one-hot vectors, can be fed directly into the AI planners without requiring ad-hoc algorithmic modifications or limiting the choice of optimization strategies.

In NIMO, the candidate pool is formulated as a single tabular dataset (see Fig. 2), where the top row serves as the header. Although the illustrated example features three experimental parameters and one objective function, NIMO inherently scales to high-dimensional spaces with arbitrary numbers of parameters and objectives simply by appending columns. Each subsequent row represents a discrete, pre-defined set of experimental conditions, comprehensively covering the intended search space. Once an experiment is executed, the quantified result is populated into the corresponding objective column. For unperformed experiments, these target entries are left blank. NIMO then determines the subsequent experimental campaign by selecting exclusively from these unmeasured entries, and these proposals are written to the proposal file (see Fig. 2). This straightforward tabular architecture functions as the universal data interface between the user and NIMO throughout the autonomous closed loop.

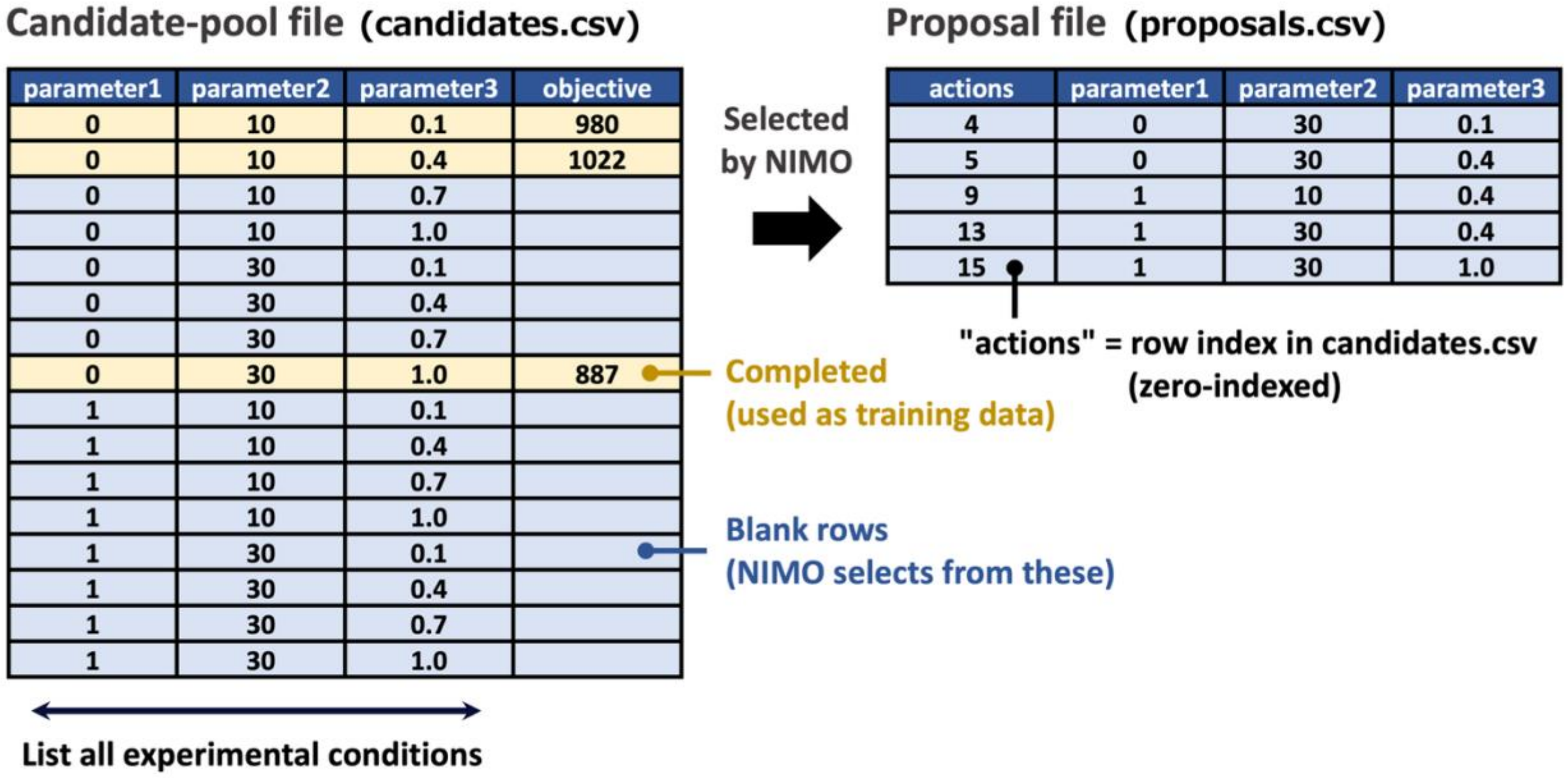

| parameter1 | parameter2 | parameter3 | objective |
|---|---|---|---|
| 0 | 10 | 0.1 | 980 |
| 0 | 10 | 0.4 | 1022 |
| 0 | 10 | 0.7 | |
| 0 | 10 | 1.0 | |
| 0 | 30 | 0.1 | |
| 0 | 30 | 0.4 | |
| 0 | 30 | 0.7 | |
| 0 | 30 | 1.0 | 887 |
| 1 | 10 | 0.1 | |
| 1 | 10 | 0.4 | |
| 1 | 10 | 0.7 | |
| 1 | 10 | 1.0 | |
| 1 | 30 | 0.1 | |
| 1 | 30 | 0.4 | |
| 1 | 30 | 0.7 | |
| 1 | 30 | 1.0 | |

| actions | parameter1 | parameter2 | parameter3 |
|---|---|---|---|
| 4 | 0 | 30 | 0.1 |
| 5 | 0 | 30 | 0.4 |
| 9 | 1 | 10 | 0.4 |
| 13 | 1 | 30 | 0.4 |
| 15 | 1 | 30 | 1.0 |

**Fig. 2** Example structure of NIMO data interfaces. (Left) The candidate-pool file (*candidates.csv*) lists all experimental conditions within the search space. Completed rows (highlighted in yellow) contain measured objective values and serve as training data, whereas rows with missing objective values (highlighted in blue) represent the unexecuted candidate pool. (Right) The proposal file (*proposals.csv*) generated by NIMO, which outputs the recommended experimental conditions. This file features an initial actions column containing zero-indexed identifiers (0, 1, 2, …) that map each proposal back to its original row index in the candidate-pool file, providing a seamless traceability link for downstream execution.

A further important consequence of the candidate-pool design is that it provides a unified architecture for diverse AI algorithms. Regardless of the underlying mathematical approach, the fundamental task required of each algorithm remains identical: to ingest the candidate pool, evaluate or rank the unmeasured entries, and export the selected conditions to the proposal file (see Fig. 1).

Consequently, every exploration algorithm in NIMO conforms to the same input/output contract and can be seamlessly invoked through a single *nimo.selection* interface, where the specific backend is determined by a single argument. This clean decoupling is precisely what makes it practical for a single software package to host the wide variety of algorithms described in Sec. 4.

## 4. AI Algorithms

NIMO ships with twelve diverse AI algorithms, each encapsulated as a self-contained module named *ai_tool_<name>.py*. Table 1 provides a side-by-side comparative summary of these algorithms, and a decision flowchart to guide algorithm selection is presented in Fig. 3. Switching between algorithms is exceptionally straightforward via the unified *nimo.selection* interface in Python, as demonstrated below:

```
nimo.selection(method = "PHYSBO",
               input_file = "candidates.csv",
               output_file = "proposals.csv",
               num_objectives = 1,
               num_proposals = 8)
```

The core optimization strategy is determined by the *method* argument (set to "PHYSBO" in this example), with certain algorithms accepting optional, method-specific parameters. The remaining arguments form a universal API common to all backends: *input_file* specifies the path to the candidate-pool dataset, *output_file* defines where the selected proposals are exported, *num_objectives* denotes the number of objective functions, and *num_proposals* dictates the batch size (the number of experimental conditions to propose concurrently). A brief technical overview of each algorithm is detailed below.

**Table 1** Summary of the twelve AI algorithms in NIMO.

| Algorithm | Category | Use case | Training data |
|---|---|---|---|
| PHYSBO | Bayesian optimization | Maximize / minimize 1–3 objectives | Required |
| BOMP | Process-constrained BO | Batch experiments with shared process parameters | Required |
| NTS | Nested Thompson sampling | Threshold-based selection with diversity option | Required |
| COMBI | Composition-spread BO | Combinatorial sputtering / spread experiments | Required |
| PTR | Target-range probability | Discover candidates inside specified property windows | Required |
| BLOX | Objective-free exploration | Diverse coverage of property space | Required |
| PDC | Active learning | Phase diagram construction (categorical labels) | Required |
| RSVM | Ranking-based learning | Ordinal optimization with optional transfer learning | Required |
| RE | Random sampling | Initial seeding, baseline | Not required |

| Algorithm | Category | Use case | Training data |
|---|---|---|---|
| | | comparison | |
| DOE | Design of experiments | Space-filling initial datasets (greedy/distance/D-opt/LHS) | Not required |
| ES | Exhaustive search | Sequential walk-through of small candidate pools | Not required |
| LLMEP | LLM-based planning | Natural-language-guided selection; supports text-valued objectives | Optional |

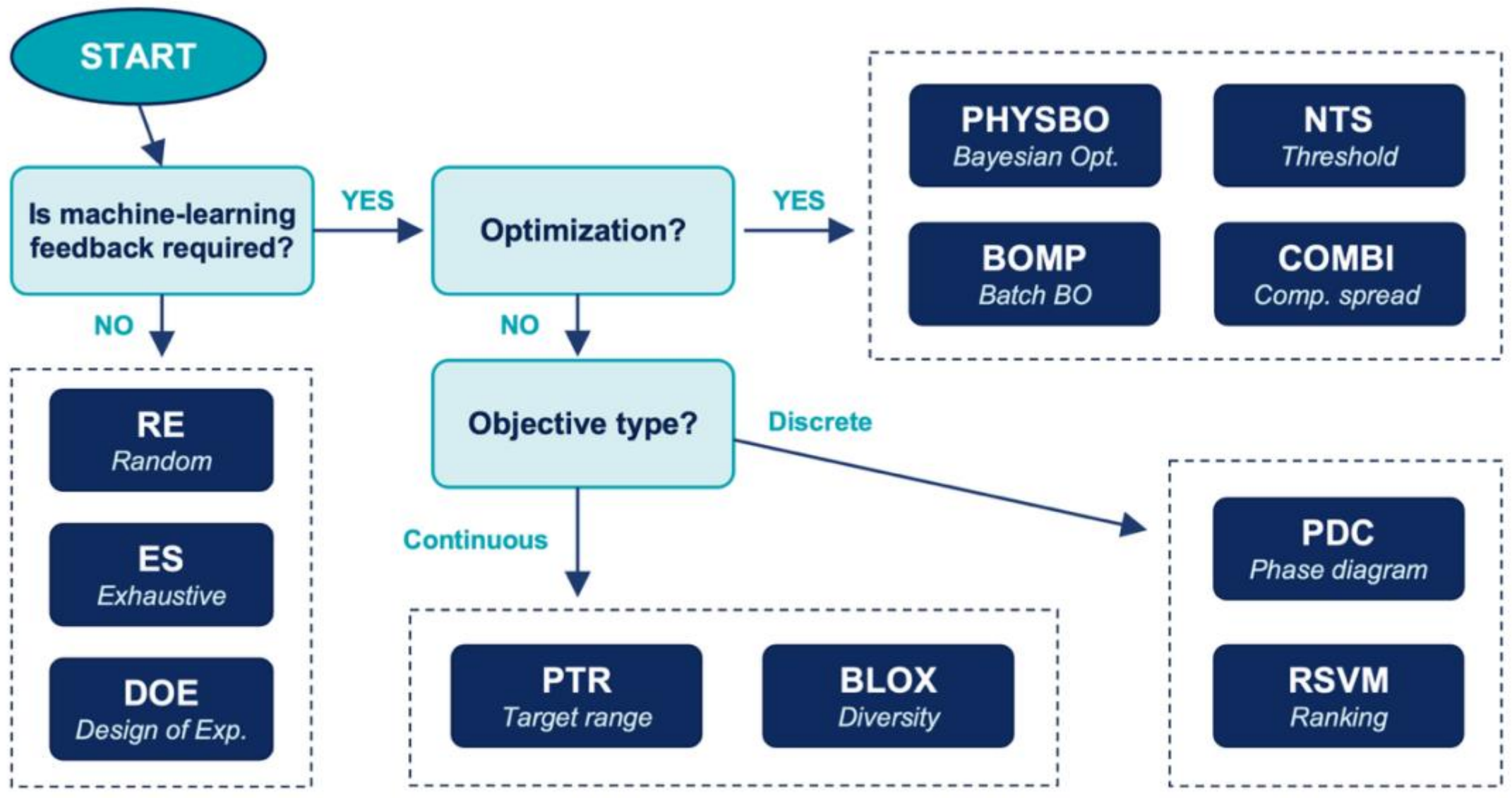

**Fig. 3** Decision-flow chart for selecting one of the core algorithms in NIMO based on the user's exploration goal.

### 4.1 Optimization

**PHYSBO.** The primary workhorse of NIMO is Bayesian optimization (BO) powered by the PHYSBO package[25]. Unlike conventional BO tools tailored for continuous search spaces, PHYSBO is uniquely optimized for discrete candidate pools. It employs Gaussian process regression to predict the objective value and predictive variance for each candidate, subsequently selecting the next experimental conditions by maximizing a designated acquisition function. For single-objective optimization, NIMO supports Thompson sampling (TS), expected improvement (EI), and probability of improvement (PI), defaulting to EI. For multi-objective optimization, it supports TS, hypervolume probability of improvement (HVPI), and expected hypervolume improvement (EHVI), defaulting to TS. Additionally, PHYSBO natively accepts a minimization flag for property minimization tasks and supports batch proposals for concurrent multi-experiment scheduling. Because a candidate-pool framework requires calculating predictions across the entire search space simultaneously, the computational overhead scales with the pool size. To mitigate this issue, PHYSBO incorporates acceleration techniques such as parallelized computing and random feature maps, which reduce execution time and enable scalable optimization even over large candidate pools.

**BOMP (Bayesian Optimization with Material Process).** Many real-world materials exploration campaigns exhibit operational constraints where certain process variables must remain fixed within a given batch. For example, a single heat-treatment temperature is inherently shared across multiple compositions deposited onto the same substrate or wafer. BOMP explicitly enforces this hierarchical structure by holding designated parameters constant during a specific batch while optimizing only the remaining free variables. This approach is substantially more sample-efficient than standard optimization routines that treat process variables as fully independent free parameters, particularly when those variables cannot be modulated at the individual sample level due to hardware limitations.

**NTS (Nested Thompson Sampling).** NTS targets scenarios where the primary objective is to identify candidates that exceed a dynamic, moving threshold defined relative to the best observed value to date[26]. To flexibly tune the balance between exploitation and exploration, the sampling mode can be selected from three pre-defined regimes: 'aggressive' (0.99 × present best), 'moderate' (0.95 × present best), and 'conservative' (0.9 × present best). By employing this threshold-centric selection strategy, NTS effectively identifies near-optimal candidate conditions while explicitly preserving sample diversity throughout the search space.

**COMBI (Composition-Spread Exploration).** COMBI is purpose-built for combinatorial composition-spread experiments[27]. The algorithm operates in three sequential phases: it first identifies a promising composition point via Bayesian optimization, then proposes a set of compositions along a continuous gradient between specified element pairs, and finally identifies the most informative element pair for the next experimental campaign. This algorithm explicitly mimics the physics of combinatorial sputtering, where a single deposition run simultaneously synthesizes a wide range of compositions on a single substrate by deliberately adjusting the target-substrate geometry.

### 4.2 Target-range and objective-free exploration

**PTR (Probability of Target Range).** In many material development campaigns, the ultimate objective is not the absolute maximization of a single property, but rather steering multiple properties into a target operating window. PTR performs Gaussian process regression on each individual objective function and subsequently proposes candidates that maximize the joint probability of falling within user-specified ranges[28]. This approach is exceptionally powerful for multi-property design criteria, for example, in a scenario where a material must simultaneously satisfy a narrow bandgap window between 1.6 and 1.8 eV and a formation energy below a designated stability threshold.

**BLOX (Boundless Objective-free eXploration).** Rather than seeking a single optimal material, BLOX aims to map out the broader property space by acquiring a diverse set of candidates as uniformly as possible[29]. It leverages the Stein discrepancy evaluated on a random-forest regression model as a metric to quantify the novelty of each unmeasured candidate relative to the existing experimental training data. To generate batch proposals concurrently, BLOX employs an iterative greedy selection strategy: it sequentially selects the candidate exhibiting the largest Stein discrepancy, treats its model-predicted value as a pseudo-observation, and updates the state before choosing the next proposal. This objective-free paradigm is uniquely suited for tasks such as constructing comprehensive materials libraries or uncovering highly anomalous outliers that traditional optimization routines would inherently overlook.

### 4.3 Categorical and ranking-based exploration

**PDC (Phase Diagram Construction).** PDC fundamentally differs from the aforementioned algorithms in that the target objective is a categorical phase label rather than a continuous property[30–32]. It employs graph-based semi-supervised classification models, specifically, either Label Propagation ('LP') or Label Spreading ('LS'), to infer the phase labels across all unmeasured candidate conditions. Subsequently, PDC applies an active-learning uncertainty sampling strategy, offering options such as Least Confidence ('LC'), Margin Sampling ('MS'), or Entropy ('EA'), to propose the next experimental conditions that maximally reduce the classification uncertainty near phase boundaries.

**RSVM (Ranking Support Vector Machine).** RSVM operates on ordinal relationships rather than absolute values: instead of directly regressing each objective function, it learns the relative ranking among observed samples to predict the preference order of unmeasured candidates[28]. This formulation is exceptionally robust when absolute objective values are plagued by high noise or scale uncertainties, yet their relative ordering remains reliable. Furthermore, RSVM can seamlessly incorporate ranking data from related but distinct material systems as auxiliary datasets for transfer learning, making it highly effective when leveraging prior knowledge from a closely aligned historical campaign.

### 4.4 Initial sampling without training data

The subsequent three algorithms belong to a distinct subgroup of initial-sampling methods; they require no prior training data and are typically deployed either to seed the inaugural cycle of an autonomous closed loop or to serve as baseline controls for comparison with more sophisticated exploration frameworks.

**RE (Random Exploration).** RE selects candidates uniformly at random from the unmeasured pool. This method is primarily useful for generating unbiased initial training datasets or acting as a baseline control to quantify the relative performance gains of advanced algorithms.

**DOE (Design of Experiments).** DOE implements four classical sampling paradigms: greedy, distance-based, D-optimal, and Latin hypercube sampling (LHS). Compared to RE, DOE algorithms produce structured, space-filling initial datasets designed to cover the available candidate pool with significantly greater uniformity.

**ES (Exhaustive Search).** ES systematically iterates through the candidate pool in sequential row order. This approach is most advantageous when the entire search space is compact enough to be fully characterized, or when the user explicitly intends to enforce a strict, deterministic experimental sequence.

### 4.5 LLM-based experimental planning

**LLMEP (LLM-based Experimental Planner).** LLMEP delegates the candidate selection step to a large language model (LLM) rather than a traditional numerical optimizer[33]. The system constructs a comprehensive Markdown prompt encompassing the experimental context, such as the target chemistry, optimization objectives, tabulated prior measurements, and highly nuanced domain constraints, and pairs it with a system prompt that defines the LLM's expert scientific persona. LLMEP is uniquely advantageous for complex exploration tasks where the target objective cannot be reduced to a single numerical value. It natively processes phase names, crystal structures, and free-text qualitative descriptors within the objective column, enabling the LLM to reason over multi-modal or text-based data that conventional regression-based optimizers inherently fail to represent. Importantly, while LLMEP can conceptually encompass the functionalities of the aforementioned algorithms, utilizing a dedicated algorithm is highly recommended whenever the experimental goal aligns with a specific mathematical framework; for instance, PHYSBO remains the preferred choice for standard Bayesian optimization tasks.

## 5. Self-Driving Laboratory Implementations

We have validated NIMO's modular architecture across multiple distinct SDL implementations spanning diverse experimental modalities (see Fig. 4). Crucially, all these workflows share the exact same *nimo.selection* core execution logic and identical candidate file formats; the integration between NIMO and the respective robotic hardware platforms was achieved via lightweight, project-specific wrapper code. In the following subsections, we outline six primary demonstrations, NAREE, CHEMSPEED, COMBAT, ROPES, the Coffee Ring SDL, and BioDot, which collectively illustrate NIMO's exceptional versatility across chemical, material, and biological domains. Beyond these primary setups, NIMO's core capability has also been successfully integrated into other automated discovery frameworks, such as the BRINE system[34].

### 5.1 NAREE: Autonomous electrochemistry

NAREE (NIMS Automated Robotic Electrochemical Experiments) [35,36], is a high-throughput, microplate-based electrochemical platform designed to automate both electrolyte formulation and batter performance evaluation. The hardware orchestrates a liquid-handling dispenser to mix chemical solutions according to recipes, a robotic arm to transport multi-well plates between functional stations, and a parallelized electrochemical measurement unit capable of characterizing up to 96 cells concurrently. By integrating NIMO, the NAREE platform was successfully closed into a fully autonomous and automated electrolyte-discovery loop[23]. As a benchmark demonstration, we targeted the exploration of multi-component electrolyte formulations for lithium-metal anodes, selecting 5 additives from a pool of 16 candidates to evaluate a combinatorial space of 4,368 unique combinations. Utilizing the discharge time as the objective function within a fully autonomous closed loop, NIMO efficiently navigated this massive formulation space and successfully identified the optimal electrolyte composition.

### 5.2 CHEMSPEED: Automated chemical synthesis

A Chemspeed automated synthesis platform is seamlessly coupled with supercritical fluid chromatography (SFC) to establish an end-to-end chemical reaction and characterization workflow[37]. By integrating NIMO, we equipped this high-throughput hardware infrastructure with BBO capabilities. The autonomous loop operates via a streamlined data pipeline: SFC analytical results are automatically parsed to extract reaction yields and selectivities, NIMO invokes its BBO module to propose the next reaction parameters, and these newly generated conditions are dispatched back to Chemspeed system for the subsequent execution cycle. Through this closed-loop exploration, we targeted the optimization of a Suzuki-Miyaura coupling reaction. NIMO successfully navigated the discrete multi-variable space, encompassing various ligand types, bases, and solvent systems, steadily enhancing the yield of the target molecule and rapidly converging on the optimal reaction conditions.

### 5.3 COMBAT: Composition-spread thin-film synthesis

The COMBAT system pairs combinatorial sputtering with the COMBI algorithm to execute autonomous thin-film material exploration[27,38]. The experimental workflow comprises sequential phases: composition-spread film deposition via combinatorial sputtering, photoresist-free fabrication of a 13-device array via laser patterning, and simultaneous anomalous Hall effect (AHE) measurements across all 13 integrated devices using a customized multichannel probe. Because samples are transferred between these functional stations manually, the overall campaign operates in a well-orchestrated human-in-the-loop manner. In a recent benchmark demonstration powered by NIMO, the COMBAT system optimized five-element (quinary) alloy compositions to maximize the AHE signal, a key figure of merit for next-generation spintronic devices, within an autonomous, closed-loop exploration framework.

### 5.4 ROPES: Automated process informatics for fuel cells

The ROPES (Robotic Objective Process Exploration System) framework focuses on process informatics (PI) to optimize the manufacturing sequences of functional materials[39]. Specifically, the "Coating and Drying ROPES" setup serves as a miniature pilot line that emulates industrial-scale fuel cell catalyst layer manufacturing. The hardware infrastructure orchestrates a slot-die coater for precise catalyst ink deposition, an automated hot-air drying oven for catalyst membrane solidification, and integrated characterization tools to evaluate film quality. By linking this automated pilot line with NIMO, the system can autonomously explore and refine highly complex operational parameters. This NIMO-driven closed-loop manufacturing pipeline is designed to circumvent traditional trial-and-error workflows.

### 5.5 Coffee Ring SDL

The coffee ring SDL[24] was developed to automate the study of the coffee ring effect — the ring-like residue left behind when a particle-laden droplet dries on a surface. The SDL combines a robotic arm, an OT-2 liquid handler (opentrons), a hotplate, and a camera. It prepares silica solutions containing two surfactants (PVA: Polyvinyl alcohol and DTAB: Dodecyltrimethylammonium bromide), heats the droplets, and classifies the resulting pattern as "ring" or "no ring" using a custom image-analysis pipeline. This SDL also served as the demonstration platform for NIMO's integration with IvoryOS (see Sec. 2): NIMO's PDC algorithm actively sampled experimental conditions near the boundary between "ring" and "no ring", while IvoryOS provided the web-based interface for designing and running the campaign. The campaign successfully mapped the coffee-ring phase diagram in the PVA–DTAB concentration space, confirming that NIMO's algorithms can drive an IvoryOS-managed campaign without any modification.

### 5.6 BioDot: Legacy liquid-handler integration

The BioDot AD1520 is a contactless precision dispenser whose native control environment relies on a legacy Visual Basic (VB) program running on a Windows platform. This setup exemplifies the exact type of legacy, non-Python-native automation that conventional Python-centric self-driving laboratory (SDL) frameworks routinely struggle to incorporate. We seamlessly integrated this BioDot system into an autonomous closed-loop SDL by exploiting NIMO's highly flexible file-exchange interface[40]. Specifically, NIMO writes the continuous or discrete dispensing parameters determined by its AI module to a shared file, which the existing VB program automatically reads to orchestrate the hardware execution. This successful deployment demonstrates that NIMO can seamlessly absorb VB-controlled or LabVIEW-controlled instruments without requiring any modification to the legacy software stack. This capability is of substantial practical importance, as it offers a cost-effective, non-disruptive pathway to upgrade existing laboratory infrastructure into modern autonomous discovery loops.

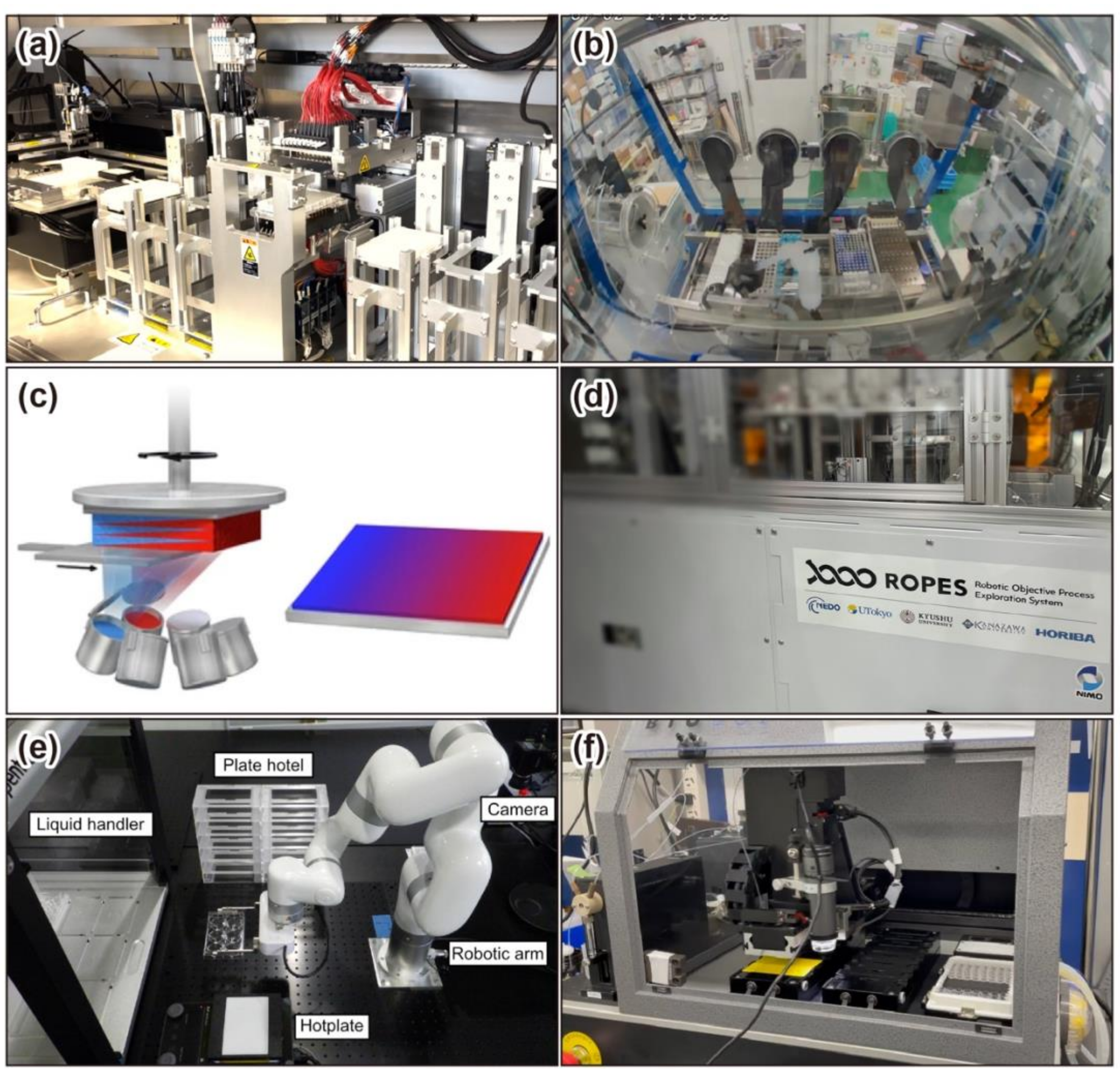

**Fig. 4** NIMO-integrated SDL implementations. (a) NAREE: a microplate-based electrochemical platform that automates electrolyte preparation and battery-performance evaluation[23]. (b) Chemspeed: an automated synthesis platform coupled with supercritical fluid chromatography (SFC) for end-to-end chemical-reaction and characterization workflows[37]. (c) Schematic illustration of COMBAT: a combinatorial sputtering system that fabricates composition-spread thin films and characterizes them via parallel anomalous Hall effect measurements on a single substrate[27]. (d) ROPES: a robotic pilot-line platform that emulates industrial fuel-cell catalyst-layer manufacturing through automated coating and drying operations[39]. (e) Coffee Ring SDL: an autonomous setup combining a robotic arm, a liquid handler, a hotplate, and a camera to map the phase boundary of the coffee-ring effect, which also serves as the demonstration platform for NIMO–IvoryOS interoperability[24]. (f) BioDot AD1520: a contactless precision dispenser controlled by a legacy Visual Basic program, integrated through NIMO's file-exchange interface[40].

## 6. NIMO Desktop

While NIMO's Python interface is ideal for developing fully automated self-driving laboratories (SDLs), it may pose a barrier in human-in-the-loop scenarios, as many experimental scientists do not code daily. NIMO features a native cross-platform desktop application, NIMO Desktop (available for Windows and macOS at https://www.nims.go.jp/nimo/).

As illustrated in Fig. 5, NIMO Desktop features a two-tab interface that streamlines the closed-loop exploration workflow described in Sec. 2.

**Selection tab.** This tab enables users to generate subsequent experimental proposals (Fig. 5(a)). The

workflow involves:

(i) Selecting one of nine built-in algorithms (RE, ES, DOE, PHYSBO, BOMP, NTS, PTR, BLOX, PDC) by clicking the corresponding button

(ii) Specifying the input candidate-pool file and output proposal file

(iii) Defining the number of objectives and proposals

Crucially, the candidate-pool file format is identical to that of the Python API (see Sec. 3), ensuring seamless data compatibility. Selecting certain algorithms dynamically reveals specific parameter inputs; for instance, PTR prompts for a target range, BOMP requires fixed parameters across the batch, and NTS allows selection of the sampling mode (conservative, moderate, or aggressive). Clicking "Start Calculation" executes the algorithm and exports the proposals.

**Update tab.** This tab allows users to log experimental outcomes back into the candidate-pool file (Fig. 5(b)). Upon loading the proposal file, the GUI dynamically generates input fields for each proposal, allowing direct entry of measured objectives. Unsuccessful or incomplete experiments can simply be left blank, which leaves the corresponding objective values blank in the candidate-pool file. Clicking "Update Candidates" commits these values to the file.

By cycling between these tabs, users maintain an active-learning loop that incrementally aggregates data. NIMO Desktop thus democratizes advanced experimental optimization, making it fully accessible to non-programming scientists while sharing the exact same data structures as the developer-focused Python API.

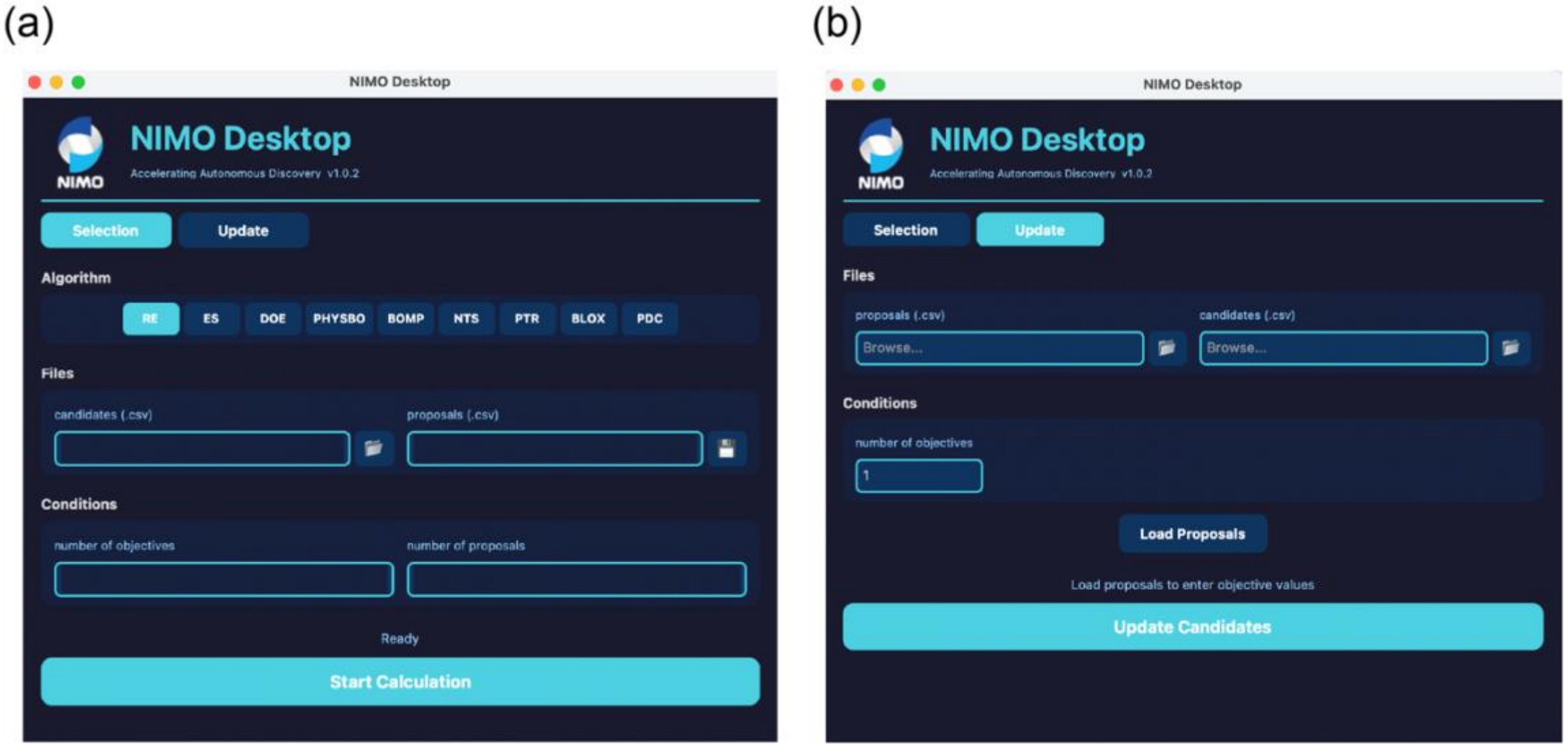

**Fig. 5** Screenshots of NIMO Desktop interface. (a) Selection tab: Select an algorithm, specify the paths of the candidate pool and proposal files, set the number of objectives and proposals, and click “Start Calculation”. (b) Update tab: After performing the proposed experiments, select the proposal file path, enter the measured objective values, and click “Update Candidates” to save these values back into the candidate-pool file.

## 7. Conclusion and Future Perspective

We have presented NIMO, a software platform that bridges the gap between AI algorithms and diverse robotic hardware in materials-science SDLs. NIMO is characterized by a modular AI-robot separation linked by a thin file-exchange contract; a discrete candidate pool design that absorbs domain knowledge and constraints as data rather than code; and a unified Python interface that exposes a portfolio of twelve AI algorithms. Notably, because these diverse AI algorithms are built in, once NIMO is integrated into an SDL, the exact same system can be readily repurposed to address materials-exploration tasks with varying objectives. The six SDL case studies (NAREE, CHEMSPEED, COMBAT, ROPES, the Coffee Ring SDL, and BioDot) demonstrate that the platform seamlessly operates across diverse domains, even when interfacing with legacy, non-Python instrumentation. Finally, the NIMO Desktop application extends this accessibility to non-programming users.

Finally, we provide a future outlook for further enhancing the capabilities of NIMO.

**Expansion of the algorithm portfolio:** A defining strength of NIMO is the breadth of its algorithm library. Future efforts will focus on maintaining this leadership through continuous algorithmic research. Planned additions include multi-fidelity Bayesian optimization, which integrates simulation and experimental data, and LLM-augmented optimizers that incorporate LLM knowledge as prior information or critique. Each new algorithm will be released as part of the open-source NIMO core.

**Enhancement of User Interface:** NIMO was originally distributed as a Python library, requiring users to write scripts to operate it. To broaden its accessibility, we have since developed multiple user interfaces tailored to different usage scenarios. NIMO Desktop provides a graphical interface designed for human-in-the-loop exploration, allowing researchers without programming experience to run closed-loop campaigns. The integration with IvoryOS, described in Sec. 2, enables workflow design through a drag-and-drop web interface, well suited to SDLs with complex experimental procedures. Most recently, we proposed NIMO Controller[41], which offers a unified interface for both human scientists and AI agents through the Model Context Protocol (a recently introduced standard for connecting AI assistants to external tools) and supports block-based workflow design. Future development will further expand this portfolio of user interfaces to support an even wider range of scenarios, from individual researchers to fully autonomous AI-driven laboratories.

**Establishment of Connected SDLs:** We envision a future in which NIMO-equipped laboratories worldwide interconnect to run collaborative materials-exploration campaigns. As a concrete example, in a multi-objective optimization task, Laboratory A could evaluate Objective 1 while Laboratory B evaluates Objective 2, with each facility contributing measurements aligned with its specific equipment and expertise. Such collaboration is naturally supported by NIMO's candidate-pool design (Sec. 3). Because the only data shared between laboratories is a single candidate-pool file serving as a common reference frame, collaborative campaigns require minimal communication, regardless of the underlying instrumentation or programming environment of each laboratory.

We believe that the combination of NIMO as an AI algorithm library with interoperable orchestration (e.g., IvoryOS-style web interfaces) and LLM-mediated natural-language control will reshape laboratory automation in the near future. To realize this vision, developing AI algorithms specifically tailored for self-driving laboratories remains essential. Beyond general-purpose AI, what is truly needed to advance autonomous discovery is “AI for Science”: algorithms that integrate seamlessly with experimental workflows and natively embody the practical requirements of materials exploration.

## Author contributions

**Ryo Tamura:** Conceptualization, Methodology, Software, Writing – original draft, Writing – review & editing, Visualization, Supervision, Project administration, Funding acquisition. **Naruki Yoshikawa:** Methodology, Software, Writing – review & editing. **Koji Tsuda:** Conceptualization, Methodology, Supervision, Funding acquisition. **Shoichi Matsuda:** Conceptualization, Methodology, Writing – review & editing, Project administration, Funding acquisition.

## Conflicts of interest

There are no conflicts to declare.

## Data availability

The NIMO source code is available at https://github.com/NIMS-DA/nimo. The official documentation, including API references and worked examples, is available at https://nims-da.github.io/nimo/en/. NIMO Desktop, the no-code desktop application, is available for download at https://www.nims.go.jp/nimo/. No new experimental data were generated for this paper.

## Acknowledgments

The authors thank K. Terayama, S. Muroga, R. Shibukawa, Y. Nagata, S. Akiyama, H. Taketa, S. Murata, D. Ryuno, T. Yokota, R. Toyama, Y. Iwasaki, Y. Sakuraba, T. Ozawa, K. Nakamura, S. Muroga, K. Nagato, and W. Zhang. This work was supported by JST PRESTO (Grant No. JPMJPR24T8), JSPS KAKENHI (Grant No. JP25K21333) and the MEXT Program: Data Creation and Utilization-Type Material Research and Development Project (Digital Transformation Initiative for Green Energy Materials; Grant No. JPMXP1122712807).

## References

1 R. Ramprasad, R. Batra, G. Pilania, A. Mannodi-Kanakkithodi and C. Kim, *npj Comput. Mater.*, 2017, **3**, 1–13.
2 T. Lookman, P. V. Balachandran, D. Xue and R. Yuan, *npj Comput. Mater.*, 2019, **5**, 1–17.
3 B. P. MacLeod, F. G. L. Parlane, T. D. Morrissey, F. Häse, L. M. Roch, K. E. Dettelbach, R. Moreira, L. P. E. Yunker, M. B. Rooney, J. R. Deeth, V. Lai, G. J. Ng, H. Situ, R. H. Zhang, M. S. Elliott, T. H. Haley, D. J. Dvorak, A. Aspuru-Guzik, J. E. Hein and C. P. Berlinguette, *Sci. Adv.*, 2020, **6**, eaaz8867.
4 G. Tom, S. P. Schmid, S. G. Baird, Y. Cao, K. Darvish, H. Hao, S. Lo, S. Pablo-García, E. M. Rajaonson, M. Skreta, N. Yoshikawa, S. Corapi, G. D. Akkoc, F. Strieth-Kalthoff, M. Seifrid and A. Aspuru-Guzik, *Chem. Rev.*, 2024, **124**, 9633–9732.
5 N. Yoshikawa, Y. Asano, D. N. Futaba, K. Harada, T. Hitosugi, G. N. Kanda, S. Matsuda, Y. Nagata, K. Nagato, M. Naito, T. Natsume, K. Nishio, K. Ono, H. Ozaki, W. Shin, J. Shiomi, K. Shizume, K. Takahashi, S. Takeda, I. Takeuchi, R. Tamura, K. Tsuda and Y. Ushiku, *Digit. Discov.*, 2025, **4**, 1384–1403.
6 J. Li, C. Ding, D. Liu, L. Chen and J. Jiang, *Digit. Discov.*, 2025, **4**, 1672–1684.
7 J. Hwang, S. Kim, S. Lim, J. Kim, S. Lee, S. Min, J. Song, J. Lim, S. Hong, J.-H. Hwang, Y.-S. Choi, D.-H. Seo, S. S. Han, K. Kim, S.-H. Yoo, J. Shin, J. W. Choi, J. Nam, J. Park, J. Ryu and Y. Jung, *Digit. Discov.*, 2026, **5**, 1968–1980.
8 A. Jain, S. P. Ong, W. Chen, B. Medasani, X. Qu, M. Kocher, M. Brafman, G. Petretto, G.-M. Rignanese, G. Hautier, D. Gunter and K. A. Persson, *Concurr. Comput. Pract. Exp.*, 2015, **27**, 5037–5059.
9 D. Allan, T. Caswell, S. Campbell and M. Rakitin, *Synchrotron Radiation News*, 2019, **32**, 19–22.
10 L. M. Roch, F. Häse, C. Kreisbeck, T. Tamayo-Mendoza, L. P. E. Yunker, J. E. Hein and A. Aspuru-Guzik, *PLOS ONE*, 2020, **15**, e0229862.
11 J. Li, Y. Tu, R. Liu, Y. Lu and X. Zhu, *Adv. Sci.*, 2020, **7**, 1901957.
12 J. R. Deneault, J. Chang, J. Myung, D. Hooper, A. Armstrong, M. Pitt and B. Maruyama, *MRS Bulletin*, 2021, **46**, 566–575.
13 F. Rahmanian, J. Flowers, D. Guevarra, M. Richter, M. Fichtner, P. Donnely, J. M. Gregoire and H. S. Stein, *Adv. Mater. Interfaces*, 2022, **9**, 2101987.
14 Y. Fei, B. Rendy, R. Kumar, O. Dartsi, H. P. Sahasrabuddhe, M. J. McDermott, Z. Wang, N. J. Szymanski, L. N. Walters, D. Milsted, Y. Zeng, A. Jain and G. Ceder, *Digit. Discov.*, 2024, **3**, 2275–2288.
15 M. Sim, M. G. Vakili, F. Strieth-Kalthoff, H. Hao, R. J. Hickman, S. Miret, S. Pablo-García and A. Aspuru-Guzik, *Matter*, 2024, **7**, 2959–2977.
16 H. J. Yoo, K.-Y. Lee, D. Kim and S. S. Han, *Nat. Commun.*, 2024, **15**, 9669.
17 M. Vogler, S. K. Steensen, F. F. Ramírez, L. Merker, J. Busk, J. M. Carlsson, L. H. Rieger, B. Zhang, F. Liot, G. Pizzi, F. Hanke, E. Flores, H. Hajiyani, S. Fuchs, A. Sanin, M. Gaberšček, I. E. Castelli, S. Clark, T. Vegge, A. Bhowmik and H. S. Stein, *Adv. Energy Mater.*, 2024, **14**, 2403263.
18 M. Zaki, C. Prinz and B. Ruehle, *ACS Nano*, 2025, **19**, 9029–9041.
19 J. Gao, J. Chang, H. Que, Y. Xiong, S. Zhang, X. Qi, Z. Liu, J.-J. Wang, Q. Ding, X. Li, Z. Pan, Q. Xie, Z. Yan, J. Yan and L. Zhang, *arXiv*, 2025, preprint, arXiv:2512.21766.
20 W. Zhang, L. Hao, V. Lai, R. Corkery, J. Jessiman, J. Zhang, J. Liu, Y. Sato, M. Politi, M. E. Reish, R. Greenwood, N. Depner, J. Min, R. El-khawaldeh, P. Prieto, E. Trushina and J. E. Hein, *Nat. Commun.*, 2025, **16**, 5182.
21 K. Terayama, M. Sumita, R. Tamura and K. Tsuda, *Acc. Chem. Res.*, 2021, **54**, 1334–1346.

22 F. Häse, L. M. Roch, C. Kreisbeck and A. Aspuru-Guzik, *ACS Cent. Sci.*, 2018, **4**, 1134–1145.
23 R. Tamura, K. Tsuda and S. Matsuda, *Sci. Technol. Adv. Mater. Meth.*, 2023, **3**, 2232297.
24 N. Yoshikawa, T. Ozawa, W. Zhang, J. E. Hein, R. Tamura and S. Matsuda, *ChemRxiv*, **2026**, 15000670.
25 Y. Motoyama, R. Tamura, K. Yoshimi, K. Terayama, T. Ueno and K. Tsuda, *Comput. Phys. Commun.*, 2022, **278**, 108405.
26 R. Shibukawa, S. Matsuda, K. Nakamura, R. Tamura and K. Tsuda, *npj Comput. Mater.*, 2026, **12**, 197.
27 R. Toyama, R. Tamura, S. Matsuda, Y. Iwasaki and Y. Sakuraba, *npj Comput. Mater.*, 2025, **11**, 329.
28 W. Yuan, Y. Hibi, R. Tamura, M. Sumita, Y. Nakamura, M. Naito and K. Tsuda, *Patterns*, 2023, **4**, 100846.
29 K. Terayama, M. Sumita, R. Tamura, D. T. Payne, M. K. Chahal, S. Ishihara and K. Tsuda, *Chem. Sci.*, 2020, **11**, 5959–5968.
30 K. Terayama, R. Tamura, Y. Nose, H. Hiramatsu, H. Hosono, Y. Okuno and K. Tsuda, *Phys. Rev. Mater.*, 2019, **3**, 033802.
31 R. Tamura, G. Deffrennes, K. Han, T. Abe, H. Morito, Y. Nakamura, M. Naito, R. Katsube, Y. Nose and K. Terayama, *Sci. Technol. Adv. Mater. Meth.*, 2022, **2**, 153–161.
32 P. Zou, R. Tamura and K. Tsuda, *Digit. Discov.*, 2026, **5**, 1252–1256.
33 R. Tamura, H. Morito, Y. Oikawa, G. Deffrennes, S. Matsuda, N. Yoshikawa, T. Takayama, T. Abe, K. Tsuda and K. Terayama, *arXiv*, 2026, preprint, arXiv:2604.20304.
34 M. Ramezani, P. Nandi, P. A. D. L. Fuente-Moreno and M. Beidaghi, *Digit. Discov.*, 2026, **5**, 397–406.
35 S. Matsuda, K. Nishioka and S. Nakanishi, *Sci. Rep.*, 2019, **9**, 6211.
36 S. Matsuda, G. Lambard and K. Sodeyama, *Cell Rep. Phys. Sci.*, 2022, **3**, 100832.
37 S. Akiyama, H. Akitsu, R. Tamura, W. Matsuoka, S. Maeda, K. Tsuda and Y. Nagata, *ChemRxiv*, 2024, preprint, ChemRxiv:2024-bnj6p-v2.
38 R. Toyama, Y. Iwasaki, P. D. Kulkarni, H. Suto, T. Nakatani and Y. Sakuraba, *npj Comput. Mater.*, 2025, **11**, 269.
39 Development of an Automated Experimentation and Autonomous Exploration System for Fuel Cell Manufacturing Process, https://www.t.u-tokyo.ac.jp/en/press/pr2025-06-12-002, (accessed 1 June 2026).
40 R. Tamura, H. Taketa, S. Murata, D. Ryuno, T. Yokota, K. Tsuda and S. Matsuda, *Sci. Technol. Adv. Mater. Meth.*, 2025, **5**, 2565144.
41 N. Yoshikawa and R. Tamura, *arXiv*, 2026, preprint, arXiv:2605.15227.